\documentclass{ws-mpla}
\usepackage{cite,slashed}
\usepackage{graphicx,graphbox,xcolor}
\usepackage{hyperref}

\renewcommand{\d}{\mathrm{d}}

\begin{document}

\markboth{F. Z. Bara, S. Zaiem and Y. Delenda}{Top-quark pair production in $e^+e^-$ collisions within the mNCSM}

\catchline{}{}{}{}{}

\title{Top-quark pair production in electron-positron collisions within the minimal noncommutative Standard Model}

\author{Fatma Zohra Bara$^{1,2\dag}$, Slimane Zaiem$^{2\ddag*}$\footnotetext{$^*$Corresponding author.}, and Yazid Delenda$^{1,2\P}$}
\address{$^{1}$Laboratoire de Physique des Rayonnements et de leurs Interactions avec la Mati\`{e}re,\\
$^{2}$D\'{e}partement de Physique, Facult\'{e} des Sciences de la Mati\`{e}re,\\
Universit\'{e} de Batna-1, Batna 05000, Algeria\\
$^\dag$fatmazohra.bara@univ-batna.dz \\
$^\ddag$slimane.zaiem@univ-batna.dz\\
$^\P$yazid.delenda@univ-batna.dz}

\maketitle


\begin{abstract}
We study top-quark pair production in electron-positron collisions within the framework of the minimal noncommutative Standard Model. Noncommutative effects are incorporated using the Seiberg-Witten map, and the scattering squared amplitude for the process $e^+e^-\to t\bar{t}$ is computed consistently up to second order in the noncommutativity parameter $\Theta^{\mu\nu}$. We derive the total cross-section, the polar and azimuthal angular distributions, and the forward-backward asymmetry, all of which exhibit sensitivity to space-time noncommutativity. Numerical results are evaluated for center-of-mass energies relevant to future linear colliders, such as the ILC and CLIC. Our analysis demonstrates that noncommutative geometry can induce significant characteristic deviations from the Standard Model predictions, offering a potential indirect probe of space-time noncommutativity at high-energy $e^+e^-$ collisions.
\end{abstract}

\ccode{PACS Nos.: 11.10.Nx, 14.65.Ha, 13.66.Jn, 12.60.-i}

\section{Introduction}	

The discovery of the Higgs boson at the Large Hadron Collider (LHC) \cite{ATLAS:2012yve,CMS:2012qbp} marked a triumph of the Standard Model (SM), yet several fundamental questions remain unanswered. Among these are the nature of electroweak symmetry breaking, the origin of the observed fermion mass hierarchy, and the notably large mass of the top quark. These unresolved issues indicate that the SM is not a complete theory and motivate the search for physics Beyond the Standard Model (BSM), which has become a central focus of high-energy physics. Promising theoretical frameworks include supersymmetry \cite{Fayet:1974pd, Zaim:2008zz}, extra dimensions \cite{Antoniadis:1990ew,Arkani-Hamed:1998jmv,Antoniadis:1998ig,Appelquist:2000nn}, and noncommutative geometry \cite{Connes:1994yd}, each predicting new phenomena that could be accessible at current and future colliders. 

The top quark, with a mass near the electroweak scale, serves as a unique probe for BSM physics due to its strong coupling to the Higgs sector and potential sensitivity to new interactions. Precision studies of top-quark production and decay are therefore central to the physics programs of present and future colliders. Electron-positron colliders offer a particularly clean environment for such studies, with well-defined initial states and reduced QCD backgrounds compared to hadron colliders. Future facilities such as the International Linear Collider (ILC) \cite{Abe:2025yur,LinearColliderVision:2025hlt}, the Compact Linear Collider (CLIC) \cite{Adli:2025swq}, and others like the FCC-ee \cite{FCC:2025lpp} and CEPC \cite{CEPCStudyGroup:2023quu}, are expected to operate at energies well above the top-pair threshold and reaching the TeV scale, enabling high-precision measurements of cross-sections, angular distributions, and asymmetries as well as searches for new physics signals, exploring the plethora of models that complement the SM.

Among these many BSM scenarios, Noncommutative (NC) geometry provides an intriguing extension of space-time structure, where coordinates $\hat{x}^\mu$ satisfy a non-trivial commutation relation
\begin{equation}
[\hat{x}^{\mu},\hat{x}^{\nu}]=i\,\Theta^{\mu\nu}=\frac{i}{\Lambda_{\mathrm{NC}}^2}\,C^{\mu\nu}\,,\quad (\mu,\nu = 0,1,2,3)\,,\label{eq:1}
\end{equation}
where $\Theta^{\mu\nu}$ is an antisymmetric constant tensor encoding the scale and structure of noncommutativity. The idea of quantized space-time dates back to Snyder's early work \cite{Snyder:1946qz}, and such a structure emerges naturally in certain limits of string theory \cite{Seiberg:1999vs,Seiberg:2000ms} and quantum gravity models. This idea leads to distinctive corrections to quantum field theories defined on NC space-time, and has thus sparked considerable interest in the NC nature of space-time on the TeV-scale within the particle physics community. Using the Seiberg-Witten (SW) map, Calmet et al. \cite{Calmet:2001na} constructed the minimal Noncommutative Standard Model (mNCSM) as a consistent gauge theory on noncommutative space-time. In this framework, NC corrections appear as gauge-invariant higher-dimensional operators suppressed by $\Lambda_{\mathrm{NC}}$. To avoid potential issues with unitarity and causality \cite{Seiberg:2000ms,Minwalla:1999px,Gomis:2000zz}, one typically restricts to purely space-space noncommutativity ($\Theta^{0i}=0$), though space-time components can also be considered with careful treatment.

In canonical NC field theory, ordinary products are replaced by the Moyal-Weyl star product
\begin{equation}
(f\star g)(x)=\left.\exp\left(\frac{i}{2}\,\Theta^{\mu\nu}\,\frac{\partial}{\partial x^\mu}\,\frac{\partial}{\partial y^\nu}\right)f(x)\,g(y)\right|_{y\to x}\,,
\end{equation}
which, however, introduces challenges such as UV/IR mixing, where ultraviolet and infrared dynamics do not decouple \cite{Minwalla:1999px,VanRaamsdonk:2001jd}. A more robust approach uses the SW map, which expresses NC fields as expansions in $\Theta$ of their commutative counterparts, thus preserving gauge invariance and enabling the construction of realistic models, with demonstrated progress in perturbative renormalization \cite{Bichl:2001cq,Huang:2009zzj}.

In this approach, the resulting theory can be interpreted as a gauge-invariant effective field theory valid for energies below the characteristic NC scale $\Lambda_{\mathrm{NC}}$. It should be noted, however, that phenomena such as UV/IR mixing, which appear in the canonical star-product formulation, are not automatically captured within a finite-order SW expansion. {In particular, UV/IR mixing can induce nontrivial effects that may lead to significant phenomenological constraints in certain sectors (e.g., in the hypercharge and photon sectors), even at low energies.}

{While such effects are often formally associated with loop-level dynamics and the nonlocal structure of the theory \cite{Bichl:2001cq,Buric:2006nr,Schupp:2008fs}, they are not explicitly included in the present framework. The SW-map expansion should be treated as an effective perturbative description, whose validity is restricted to observables and energy regimes where higher-order contributions and UV/IR mixing effects do not invalidate the truncation at finite order in $\Theta^{\mu\nu}$. In this work, we focus on tree-level observables computed up to $\mathcal{O}(\Theta^{2})$, following the standard approach adopted in mNCSM phenomenological studies.}

Phenomenological studies within the mNCSM have already set preliminary bounds on the NC scale $\Lambda_{\mathrm{NC}}$ \cite{Zaim:2008zz,Behr:2002wx,Fisli:2020vzt,Ghegal:2014nea, Buric:2006nr,Alboteanu:2006hh,Ohl:2010zf,Selvaganapathy:2016jrl, Selvaganapathy:2019jkm}. For instance, analyses of $ZZ/\gamma\gamma$ \cite{Alboteanu:2006hh}, $W^+W^-$ \cite{Ohl:2010zf}, and Drell–Yan processes \cite{Selvaganapathy:2016jrl} via exotic photon-gluon-gluon and $Z$-gluon-gluon interactions at the LHC have yielded limits in the range $\Lambda_{\mathrm{NC}} \gtrsim 0.4\text{--}1\,\mathrm{TeV}$. For top-quark pair production in $e^+e^-$ collisions, earlier mNCSM studies using first‑order expansions in $\Theta$ obtained {sensitivity} around $\Lambda_{\mathrm{NC}} \sim 0.1\text{--}0.2\,\mathrm{TeV}$ \cite{Ghegal:2014nea}, \footnote{In Ref.~\cite{Ghegal:2014nea}, the NC tensor was constructed from Dirac matrices, making $\Theta^{\mu\nu}$ matrix-valued in spinor space. This choice leads to non-vanishing $\mathcal{O}(\Theta)$ correction to the cross-section. In the present work, our choice of a conventional c-number tensor $\Theta^{\mu\nu}$ shifts the leading correction to $\mathcal{O}(\Theta^2)$.} while more recent analyses of angular asymmetries have pushed the sensitivity to $\Lambda_{\mathrm{NC}} \sim 1\text{--}2.8\,\mathrm{TeV}$ using polarized beams and non-minimal couplings \cite{Fisli:2020vzt,Selvaganapathy:2019jkm}.

Studies of other clean $e^+e^-$ processes, particularly muon pair production $e^+e^- \to \mu^+\mu^-$, have shown significant sensitivity to space-time noncommutativity \cite{Prakash:2010xh, Das:2011iq}. At center-of-mass energies relevant for the ILC ($\sqrt{s}=0.5$--$1$ TeV), these analyses yield {sensitivity} on the NC scale in the range $\Lambda_{\mathrm{NC}} \gtrsim 0.8$--$1.0$ TeV \cite{Prakash:2010xh}, with the azimuthal dependence of the cross section providing a particularly clean signature. More recently, laboratory-frame analyses incorporating an averaging over Earth's rotation effects have strengthened these constraints \cite{Das:2011iq}. Other complementary bounds with Earth's rotation effects were also derived from LEP data on $e^+e^- \to q \bar{q}$, which constrain both the scale and the spatial orientation of noncommutativity \cite{Haghighat:2010up}. For the $Z\gamma$ production channel $e^+e^- \to Z\gamma$, {a sensitivity to $\Lambda_{\mathrm{NC}}$ as high as $6$ TeV has} been obtained at the ILC \cite{Alboteanu:2007zz}, demonstrating that different final states probe complementary aspects of the NC parameter.

Within the broader context of searches for Lorentz violation, the mNCSM represents a specific class of Lorentz-violating theories characterized by the tensor $\Theta^{\mu\nu}$. A more general and systematic framework for parametrizing Lorentz- and CPT-violating effects is provided by the Standard Model Extension (SME), an effective field theory that incorporates all such operators while preserving gauge invariance and renormalizability \cite{Colladay:1998fq, Kostelecky:2003fs}. Within the top quark sector, the consequences of Lorentz violation have been explored in ref. \cite{Berger:2015yha}, where the experimental signatures are presented for both $t\bar{t}$ and single-top production. On the experimental side, the CMS collaboration has recently performed the first precision test of Lorentz invariance in top quark pair production at the LHC \cite{CMS:2024rcv}, placing upper limits on Lorentz-violating couplings in the range $1-8\times 10^{-3}$ at $68\%$ confidence level, improving upon previous Tevatron constraints by up to two orders of magnitude.

In this work, we extend these analyses by computing the scattering squared amplitude for $e^+e^- \to t\bar{t}$ within the mNCSM {including contributions} up to second order in $\Theta^{\mu\nu}$ using {first-order} SW map approach. The focus on top-quark production in this study is motivated by several advantages. {While it is true that the top quark plays a special role in the SM due to its large Yukawa coupling to the Higgs boson, the process considered here proceeds dominantly through $\gamma/Z$ exchange. In this context, the relevance of the top quark arises not from Higgs-mediated interactions, but rather from its role as the heaviest fermion and its sensitivity to electroweak-scale new physics in precision observables.}

First, the large top-quark mass places it near the electroweak scale, making it particularly sensitive to {potential flavor-dependent NC corrections \cite{Bernreuther:2008ju}. This enhances the impact of higher-dimensional operators, leading to potentially larger deviations from SM predictions compared to light fermion production. Second, the $t\bar{t}$ final state provides access to a rich set of kinematic and angular observables, such as spin correlations and asymmetries. These observables are particularly sensitive to the Lorentz-violating structures encoded in the NC tensor $\Theta^{\mu\nu}$.} Third, the $t\bar{t}$ final state probes both photon and $Z$-boson exchange, offering more access to the weak interaction sector of NC corrections than light-lepton production. {Moreover, since the top quark decays before hadronization, its polarization information is directly transferred to its decay products. This provides clean access to angular distributions that are especially sensitive to the directional effects induced by space-time noncommutativity, while remaining free from non-perturbative QCD uncertainties.}

These features make $e^+e^- \to t\bar{t}$ a complementary and sensitive probe of space-time noncommutativity, with characteristic signatures in angular distributions that we analyze in detail. We provide comprehensive analytical results for the total and differential cross-sections, including polar and azimuthal distributions and the forward-backward asymmetry. Special attention is given to space-time noncommutativity, which induces characteristic directional dependencies in angular observables. Numerical evaluations are performed for center-of-mass energies relevant to the ILC and CLIC, {from which the sensitivity to $\Lambda_{\mathrm{NC}}$ is estimated} and the collision threshold energy required to observe these NC effects is established. Our goal of deriving robust constraints on the noncommutativity scale $\Lambda_{\mathrm{NC}}$ aligns with broader efforts in the field to determine small new physics couplings from precision $t\bar{t}$ data at future $e^+e^-$ colliders, where methods such as the optimal observable technique have been employed to maximize sensitivity, for example, within effective field theory approaches \cite{Bhattacharya:2023mjr}.

The paper is organized as follows. In section 2, we present the squared amplitude for $e^+e^-\to t\bar{t}$ in the mNCSM. Section 3 details the kinematics and parameterization of NC deformations. Section 4 contains numerical results for total and angular distributions, as well as the forward-backward asymmetry. Section 5 summarizes our findings and outlines prospects for future collider searches.

\section{Squared amplitude for the process $e^+e^-\to\gamma/Z\to t\bar{t}$ in the mNCSM}

\begin{figure}[ht]
\centerline{\includegraphics[width=.9\textwidth]{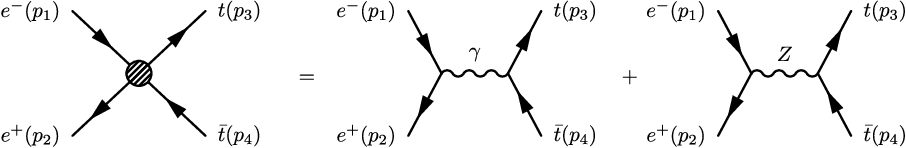}}
\vspace*{8pt}
\caption{Feynman diagrams for the process $e^-e^+ \to \gamma/Z \to t\bar{t}$ in the mNCSM.\protect\label{fig1}}
\end{figure}

We consider the pair production of top quarks in electron-positron annihilation, $e^-(p_1)e^+(p_2)\to t(p_3)\bar{t}(p_4)$, which proceeds through an $s$-channel exchange of a photon and a $Z$ boson within the framework of the mNCSM. The corresponding Feynman diagrams are shown in figure \ref{fig1}. Higgs-boson exchange is neglected, as its contribution is completely negligible. The Higgs coupling to electrons is proportional to the electron mass ($y_e \propto m_e$), leading to a strong suppression of the amplitude. Away from the Higgs pole ($s \neq m_H^2$), the propagator provides no compensating enhancement, so that the Higgs-exchange amplitude scales as $\sim m_e m_t/(s-m_H^2)$. As a result, its contribution to the cross section is suppressed by many orders of magnitude compared to $\gamma/Z$ exchange (typically $\sigma_H/\sigma_Z \lesssim 10^{-12}$ in the energy range $\sqrt{s}\sim 0.3$--$1~\mathrm{TeV}$), and can be safely neglected.

In order to consistently compute the cross section including NC effects, we employ the Feynman rules expanded up to first order in the noncommutativity parameter $\Theta_{\mu\nu}$. Details of the 
relevant Feynman rules are provided in \ref{sec:Feyn}.

It is worth noting that the SW map is not unique beyond first order in $\Theta^{\mu\nu}$. While ambiguities appearing at $\mathcal{O}(\Theta)$ cancel in on-shell observables, it has been shown that at $\mathcal{O}(\Theta^2)$ additional free parameters arise that do not, in general, cancel in physical quantities such as scattering amplitudes and cross sections \cite{Alboteanu:2007bp}. As a consequence, predictions at second order become dependent on the specific choice of SW map and are therefore less robust from a phenomenological perspective. 

A concrete illustration of these issues can be found in Ref.~\cite{Prakash:2010xh}, where the process $e^+e^- \to \mu^+\mu^-$ was studied within a \emph{non-minimal} version of the NCSM, employing Feynman rules constructed up to $\mathcal{O}(\Theta^2)$. In that framework, it was observed that the contributions to the spin-averaged squared amplitude at $\mathcal{O}(\Theta)$, $\mathcal{O}(\Theta^2)$, and $\mathcal{O}(\Theta^3)$ cancel, resulting in a leading nonvanishing correction only at $\mathcal{O}(\Theta^4)$. Such cancellations were obtained by setting all free parameters of the SW map appearing at $\mathcal{O}(\Theta^2)$ to zero, and are thus highly sensitive to the choice of these parameters entering at higher orders.

In contrast, the present work is performed within the SW at leading-order and with the minimal construction  of the NCSM, which provides a more constrained and predictive framework. For this reason, we restrict our analysis to the leading nontrivial contributions in $\Theta^{\mu\nu}$, where the theoretical predictions remain free from higher-order ambiguities and do not require additional model-dependent assumptions. This ensures that the extracted phenomenological implications are robust and directly attributable to space-time noncommutativity.

Up to $\mathcal{O}(\Theta)$, the scattering amplitudes for photon- and $Z$-boson exchange can be written in a factorized form as
\begin{subequations}
\begin{align}
i\,A_{\gamma/Z}&=i\,A_{\gamma/Z}^{\mathrm{SM}}\left(1+\frac{i}{2}\,p_2\Theta p_1\right)\left(1+\frac{i}{2}\,p_3\Theta p_4\right),\\
i\,A_{\gamma}^{\mathrm{SM}}&=-\frac{8\,\pi\,\alpha\,i}{3\,s}\left[\bar{v}(p_2)\gamma_\mu u(p_1)\right]\left[\bar{u}(p_3)\gamma^{\mu}v(p_4)\right],\\
i\,A_{Z}^{\mathrm{SM}}&=\frac{4\,\pi\,\alpha\,i}{\sin^2(2\theta_w)\left(s-m_Z^2+im_Z\Gamma_Z\right)}\left[\bar{v}(p_2)\gamma_{\mu}\left(c^e_V-c^e_{A}\gamma_5\right)u(p_1)\right]\notag\\
&\quad\times\left[\bar{u}(p_3)\gamma^{\mu}\left(c^t_V-c^t_{A}\gamma_5\right)v(p_4)\right].
\end{align}\label{eq:amplitudes}
\end{subequations}
Here we use the shorthand notation $p_i\Theta p_j \equiv p_i^\mu \Theta_{\mu\nu} p_j^\nu$, $s=(p_1+p_2)^2$ denotes the squared centre-of-mass energy, and $\alpha=e^2/(4\pi)$ is the QED coupling constant. The vector and axial-vector couplings of a fermion $i$ are given by
\begin{equation}
c_V^i=T_3^i-2\,Q_i\sin^2\theta_w\,,\qquad c_A^i=T_3^i\,,
\end{equation}
where $T_3^i$ and $Q_i$ are the weak isospin and electric charge of the fermion, respectively, and $\theta_w$ is the Weinberg mixing angle. The quantities $m_Z$ and $\Gamma_Z$ denote the mass and total decay width of the $Z$ boson. 

After summing over final-state spins and colors and averaging over the initial-state spins, the squared amplitude for the process $e^+e^- \to t\bar{t}$ can be written in the compact form
\begin{equation}
\overline{|\mathcal{A}|^2}=\overline{|\mathcal{A}^{\mathrm{SM}}|^2}\left(1+\mathcal{N}\right),
\end{equation}
where $\mathcal{N}$ encodes the leading NC correction. To $\mathcal{O}(\Theta^2)$, this correction factor is given by
\begin{equation}
\mathcal{N}=\frac{1}{4}\,(p_2\Theta p_1)^2+\frac{1}{4}\,(p_3\Theta p_4)^2+\mathcal{O}(\Theta^4)\,.\label{eq:N}
\end{equation}
The SM contribution can be decomposed into pure photon, pure $Z$, and $\gamma$--$Z$ interference terms as
\begin{equation}
\overline{|\mathcal{A}^{\mathrm{SM}}|^2}=\overline{|\mathcal{A}_\gamma^{\mathrm{SM}}|^2}+\overline{|\mathcal{A}_Z^{\mathrm{SM}}|^2}+2\,\mathrm{Re}\left(\overline{\mathcal{A}_\gamma^{\mathrm{SM}\ast}\,\mathcal{A}_Z^{\mathrm{SM}}}\right).
\end{equation}
Explicitly, these contributions read
\begin{subequations}
\begin{align}
&\overline{|\mathcal{A}_\gamma^{\mathrm{SM}}|^2}=\frac{128}{3}\,\pi^2\alpha^2\left[\frac{t^2+u^2}{s^2}+4\,\frac{m_t^2}{s}\left(1-\frac{m_t^2}{2\,s}\right)\right],\\
&\overline{|\mathcal{A}_Z^{\mathrm{SM}}|^2}=\frac{96\,\pi^2\alpha^2}{\sin^4(2\theta_w)}\,\frac{s^2}{(s-m_Z^2)^2+m_Z^2\Gamma_Z^2}\left[4\,c_A^e\,c_A^t\,c_V^e\,c_V^t\,\frac{t-u}{s}\right.\notag\\
&\qquad\left.+\left(c_A^{e\,2}+c_V^{e\,2}\right)\left(c_A^{t\,2}+c_V^{t\,2}\right)\left(\frac{t^2+u^2}{s^2}+4\,\frac{m_t^2}{s}\left(\frac{c_V^{t\,2}}{c_V^{t\,2}+c_A^{t\,2}}-\frac{m_t^2}{2\,s}\right)\right)\right],\\
&2\,\mathrm{Re}\left(\overline{\mathcal{A}_\gamma^{\mathrm{SM}\ast}\,\mathcal{A}_Z^{\mathrm{SM}}}\right)=-\frac{128\,\pi^2\alpha^2}{\sin^2(2\theta_w)}\,\frac{s(s-m_Z^2)}{(s-m_Z^2)^2+m_Z^2\Gamma_Z^2}\notag\\
&\qquad\times\left[c_V^e c_V^t\left(\frac{t^2+u^2}{s^2}+4\,\frac{m_t^2}{s}\left(1-\frac{m_t^2}{2\,s}\right)\right)+c_A^e c_A^t\,\frac{t-u}{s}\right].
\end{align}  
\end{subequations}
Here $t=(p_1-p_3)^2$ and $u=(p_1-p_4)^2$ are the usual Mandelstam variables, and $m_t$ denotes the top-quark mass. We ignore the electron mass at high energies.

In the following sections, these results are employed to compute the total cross section as well as the angular distributions for the process $e^+e^- \to t\bar{t}$ in the mNCSM.

\section{Kinematics and choices of deformation}

In this work we present predictions for both total and differential cross sections for the process $e^+e^- \to t\bar{t}$ in NC geometry. The differential angular distribution $\d\sigma/\d\Omega$ is obtained by integrating the spin- and color-summed squared amplitude over the full phase space of the anti-top momentum $p_4$ and over the magnitude of the top-quark three-momentum $|\mathbf{p}_3|$, while keeping the polar and azimuthal angles $(\theta,\phi)$ of the top quark unintegrated. The solid-angle element is $\d\Omega=\d\cos\theta\,\d\phi$. The total cross section then follows from 
\begin{equation}
\sigma=\int_{-1}^{1} \d\cos\theta\int_0^{2\pi}\d\phi\,\frac{\d\sigma}{\d\Omega}\,.
\end{equation}

After performing the phase-space integrations and imposing on-shell conditions together with four-momentum conservation, the differential angular distribution can be expressed in terms of the squared matrix element as
\begin{equation}
\frac{\d\sigma}{\d\Omega}=\frac{1}{64\,\pi^2s}\,\sqrt{1-\frac{4\,m_t^2}{s}}\,\overline{|\mathcal{A}|^2}\,.
\end{equation}
All cross sections are evaluated in the center-of-mass frame, where the produced $t\bar{t}$ pair is back-to-back. The four-momenta of the incoming and outgoing particles are chosen as
\begin{subequations}
\begin{align}
p_1&=\frac{\sqrt{s}}{2}\left(1,  0, 0, 1 \right), \\
p_2&=\frac{\sqrt{s}}{2}\left(1,  0, 0, -1 \right), \\
p_3&=\frac{\sqrt{s}}{2}\left(1,k \sin\theta\cos\phi,k\sin\theta \sin\phi, k \cos\theta \right), \\
p_4&=\frac{\sqrt{s}}{2}\left(1,-k\sin\theta\cos\phi, -k \sin\theta \sin\phi, -k \cos\theta \right),
\end{align}
\end{subequations}
with
\begin{equation}
k=\sqrt{1-4\,\frac{m_t^2}{s}}\,.
\end{equation}
In this parametrization the Mandelstam variables reduce to
\begin{subequations}
\begin{align}
t&=m_t^2-\frac{s}{2}\left(1- k\cos\theta\right),\\
u&=m_t^2-\frac{s}{2}\left(1+ k\cos\theta\right).
\end{align}
\end{subequations}

In the NC extension of the SM, the antisymmetric tensor $\Theta^{\mu\nu}$ contains six independent parameters in four-dimensional space-time. It can be decomposed into space-time and space-space components, $\Theta^{0i}$ and $\Theta^{ij}$ ($i,j=1,2,3$), which are conveniently parametrized in terms of a characteristic NC scale $\Lambda_{\mathrm{NC}}$ and two dimensionless vectors $\boldsymbol{\xi}$ and $\boldsymbol{\beta}$ as
\begin{subequations}
\begin{align} 
\Theta^{0i}&=\frac{1}{\Lambda_{\mathrm{NC}}^2}\,\xi_i\,,\\
\Theta^{ij}&=\frac{1}{\Lambda_{\mathrm{NC}}^2}\,\epsilon_{ijk}\beta^k\,.
\end{align}
\end{subequations}
The structure of $\Theta^{\mu\nu}$ is formally analogous to that of the electromagnetic field strength tensor, with $\boldsymbol{\xi}$ and $\boldsymbol{\beta}$ playing the role of effective electric- and magnetic-like components, respectively. In the context of a simplified isotropic scenario, these parameters are set as $\xi_i = \beta_i = 1/\sqrt{3}$. Anisotropic NC space-times with variable $\xi_i$ and $\beta_i$ are equally physically admissible.

While purely space-space noncommutativity preserves unitarity and causality at the perturbative level, it leads to a strong suppression of observable effects in many $2\to2$ scattering processes. In particular, for $s$-channel reactions such as $e^+e^-\to t\bar{t}$, the leading NC corrections proportional to $(p_2\Theta p_1)$ and $(p_3\Theta p_4)$ vanish identically in the center-of-mass frame when only $\Theta^{ij}$ is nonzero. Consequently, purely space-space noncommutativity fails to generate any deviation from the SM prediction at leading order, rendering this scenario phenomenologically uninformative for collider studies for the particular process of interest in this paper.

Allowing for space-time noncommutativity ($\Theta^{0i}\neq 0$) restores nonvanishing leading-order corrections and induces characteristic polar and azimuthal modulations in angular observables. It is well known, however, that space-time noncommutativity
generically leads to an apparent violation of perturbative unitarity, manifested by a rapid growth of total cross sections with increasing center-of-mass energy. Within the framework of the mNCSM, this behavior should be interpreted as signaling the
breakdown of the effective field-theory (EFT) description near the NC scale $\Lambda_{\mathrm{NC}}$, rather than as a fundamental inconsistency of the underlying theory.

Since the mNCSM is constructed as an EFT expanded in powers of the noncommutativity parameter $\Theta\sim \Lambda_{\mathrm{NC}}^{-2}$, perturbative control requires the dimensionless expansion parameter $s/\Lambda_{\mathrm{NC}}^{2}$ not to be
parametrically large. In practice, phenomenologically relevant NC effects arise when $\Lambda_{\mathrm{NC}}$ is of the same order as the collider energy $\sqrt{s}$. For $\Lambda_{\mathrm{NC}}\gg \sqrt{s}$ the NC-induced deviations rapidly become
suppressed and phenomenologically unobservable, while for $\Lambda_{\mathrm{NC}}\ll \sqrt{s}$ the EFT description ceases to be reliable. Accordingly, meaningful phenomenological analyses should focus on the intermediate regime $\Lambda_{\mathrm{NC}}\gtrsim \sqrt{s}$ and primarily on differential observables, such as polar and azimuthal angular distributions, where NC effects arise through interference with the SM amplitudes and remain under qualitative perturbative control.

In this study we parameterize the space-time NC tensor as
\begin{equation}
\Theta^{\mu\nu}=\frac{1}{\Lambda_{\mathrm{NC}}^2}
\begin{pmatrix}
0    & \xi_x & \xi_y & \xi_z \\
-\xi_x & 0 & 0 & 0 \\
-\xi_y & 0 & 0 & 0 \\
-\xi_z & 0 & 0 & 0 
\end{pmatrix},
\end{equation}
where the dimensionless unit vector $(\xi_x,\xi_y,\xi_z)$ specifies the orientation of the space-time deformation, in analogy with an electric-field direction. The inclusion of additional space-space components does not affect our results. For a general space-time deformation, the NC correction factor introduced in eq.~\eqref{eq:N} can be written as
\begin{equation}
\mathcal{N}=\frac{s^2}{16\,\Lambda_{\mathrm{NC}}^4}\left[\xi_z^2+\left(1-4\frac{m_t^2}{s}\right)\left(\xi_z\cos\theta+\xi_x\sin\theta\cos\phi+\xi_y\sin\theta\sin\phi\right)^2\right]\,.
\end{equation}

In the following we specialize to the discussion of two representative scenarios: (i) a transverse deformation in the plane perpendicular to the beam axis, $(\xi_x,\xi_y,0)$, for which
\begin{equation}
\mathcal{N}_T=\frac{s^2}{16\,\Lambda_{\mathrm{NC}}^4}\left(1-4\,\frac{m_t^2}{s}\right)\sin^2\theta\left(\xi_x\cos\phi+\xi_y\sin\phi\right)^2\,,\label{eq:Nt}
\end{equation}
and (ii) a purely longitudinal deformation along the beam direction, $(0,0,\xi_z)$ with $\xi_z=1$, leading to
\begin{equation}
\mathcal{N}_L=\frac{s^2}{16\,\Lambda_{\mathrm{NC}}^4}\left[1+\left(1-4\,\frac{m_t^2}{s}\right)\cos^2\theta\right].
\end{equation}
These two benchmark choices capture the essential qualitative features of NC effects in the angular distributions of the $e^+e^-\to t\bar{t}$ process.

We note that in an actual experiment, the fixed noncommutativity vector $\boldsymbol{\xi}$ is constant in the Sun-centered inertial frame. As the Earth rotates, the orientation of $\boldsymbol{\xi}$ relative to the detector changes with sidereal time. Consequently, the coefficients of the azimuthal modulation in Eq.~\eqref{eq:Nt} become time-dependent, causing the cross section to exhibit sidereal-time variations.

Experimentally, this strategy has been implemented by the CMS collaboration in a search for Lorentz-invariance violation in $t\bar{t}$ production at the LHC, using the normalized differential cross section as a function of sidereal time \cite{CMS:2024rcv}. No evidence for Lorentz violation was observed; instead, the analysis was used to place upper limits on the corresponding Lorentz-violating couplings by exploiting the time dependence induced by Earth's rotation with respect to a fixed inertial frame. A similar sidereal-time analysis could in principle be applied to $e^+e^-\to t\bar{t}$ at future colliders, potentially providing greater sensitivity to the NC scale $\Lambda_{\mathrm{NC}}$ than time-averaged measurements.

\section{Total and angular differential cross-sections}

\subsection{Total cross-section}

We begin by examining the impact of NC geometry on the total cross section. After averaging over the azimuthal angle, the NC correction factor takes the form
\begin{equation}\label{eq:azim}
\int_0^{2\pi}\!\mathcal{N}\,\d\phi=\frac{s^2}{16\,\Lambda_{\mathrm{NC}}^4}\,2\pi\left[\xi_z^2+\left(1-4\,\frac{m_t^2}{s}\right)\left(\xi_z^2\cos^2\theta+\frac{1}{2}\left(\xi^2_x+\xi_y^2\right)\sin^2\theta\right)\right].
\end{equation}
Upon integration over the polar angle, the total cross section can be written as
\begin{equation}
\sigma = \sigma_{\mathrm{SM}}+\delta\sigma_{\mathrm{NC}}\,,
\end{equation}
where $\sigma_{\mathrm{SM}}$ denotes the SM contribution and $\delta\sigma_{\mathrm{NC}}$ encodes the NC correction.

The SM total cross section is given by
\begin{align}
\sigma^{\mathrm{SM}}&=\frac{4\pi\alpha^2}{s}\sqrt{1-4\,\frac{m_t^2}{s}}\,\Bigg\{\frac{4}{9}\left(1+2\,\frac{m_t^2}{s}\right)\notag\\
&\quad+\frac{c_A^{e\,2}+c_V^{e\,2}}{\sin^4(2\theta_w)}\,\frac{s^2}{\left(s-M_Z^2\right)^2+\Gamma_Z^2M_Z^2}\left[c_A^{t\,2}\left(1-4\,\frac{m_t^2}{s}\right)+c_V^{t\,2}\left(1+2\,\frac{m_t^2}{s}\right)\right]\notag\\
&\quad-\frac{4}{3}\,\frac{c_V^e\,c_V^t}{\sin^2(2\theta_w)}\,\frac{s\left(s-M_Z^2\right)}{\left(s-M_Z^2\right)^2+\Gamma_Z^2M_Z^2}\left(1+2\,\frac{m_t^2}{s}\right)\Bigg\}\,.
\end{align}
The NC contribution can be decomposed into longitudinal and transverse components,
\begin{equation}
\delta\sigma_{\mathrm{NC}}=\delta\sigma^{\mathrm{NC}}_L+\delta\sigma^{\mathrm{NC}}_T\,,
\end{equation}
corresponding to deformations parallel and perpendicular to the beam axis, respectively.

For Longitudinal space-time deformations, the NC correction reads
\begin{align}
\delta\sigma^{\mathrm{NC}}_L&=\xi_z^2\,\frac{s^2}{80\,\Lambda_{\mathrm{NC}}^4}\,\frac{4\pi\alpha^2}{s}\sqrt{1-4\,\frac{m_t^2}{s}}\,\Bigg\{\frac{4}{9}\left(7+4\,\frac{m_t^2}{s}-8\,\frac{m_t^4}{s^2}\right)\notag\\
&\quad+\frac{c_A^{e\,2}+c_V^{e\,2}}{\sin^4(2\theta_w)}\,\frac{s^2}{\left(s-M_Z^2\right)^2+\Gamma_Z^2M_Z^2} \notag\\
&\quad\times\left[7\left(c_V^{t\,2}+c_A^{t\,2}\right)+4\,\frac{m_t^2}{s}\left(c_V^{t\,2}-9\,c_A^{t\,2}\right)-8\,\frac{m_t^4}{s^2}\left(c_V^{t\,2}-4\,c_A^{t\,2}\right)\right]\notag\\
&\quad-\frac{4}{3}\,\frac{c_V^e\,c_V^t}{\sin^2(2\theta_w)}\,\frac{s\left(s-M_Z^2\right)}{\left(s-M_Z^2\right)^2+\Gamma_Z^2M_Z^2} \left(7+4\,\frac{m_t^2}{s}-8\,\frac{m_t^4}{s^2}\right)\Bigg\}\,,
\end{align} 
and for transverse space-time deformations one obtains
\begin{align}
\delta\sigma^{\mathrm{NC}}_T&=\left(\xi_x^2+\xi_y^2\right)\frac{s^2}{160\,\Lambda_{\mathrm{NC}}^4}\,\frac{4\pi\alpha^2}{s}\sqrt{1-4\frac{m_t^2}{s}}\,\Bigg\{\frac{4}{9}\left(1-4\,\frac{m_t^2}{s}\right)\left(3+8\,\frac{m_t^2}{s}\right)\notag\\
&+\frac{c_A^{e\,2}+c_V^{e\,2}}{\sin^4(2\theta_w)}\,\frac{s^2}{\left(s-M_Z^2\right)^2+\Gamma_Z^2M_Z^2} \notag\\
&\times\left(1-4\,\frac{m_t^2}{s}\right)\left[3\,c_A^{t\,2}\left(1-4\,\frac{m_t^2}{s}\right)+c_V^{t\,2}\left(3+8\,\frac{m_t^2}{s}\right)\right]\notag\\
&-\frac{4}{3}\,\frac{c_V^e\,c_V^t}{\sin^2(2\theta_w)}\,\frac{s\left(s-M_Z^2\right)}{\left(s-M_Z^2\right)^2+\Gamma_Z^2M_Z^2}\left(1-4\,\frac{m_t^2}{s}\right)\left(3+8\,\frac{m_t^2}{s}\right)\Bigg\}\,.
\end{align} 

\begin{figure}[ht]
\centerline{\includegraphics[width=.45\textwidth]{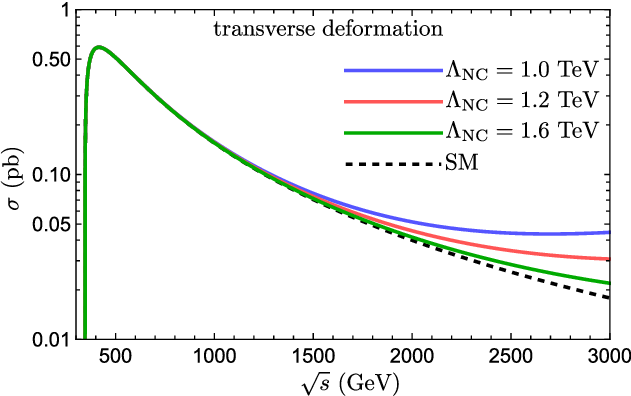}
\includegraphics[width=.45\textwidth]{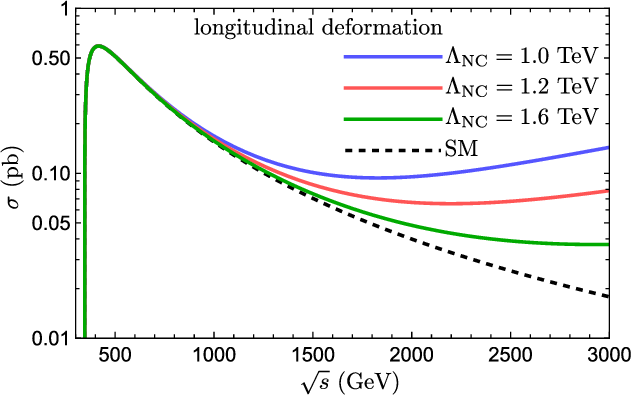}}
\vspace*{8pt}
\caption{\label{fig:total}Total cross-section for transverse (left) and longitudinal (right) space-time deformations as a function of the center-of-mass energy $\sqrt{s}$.\protect}
\end{figure} 

In figure \ref{fig:total} we display the total cross section as a function of the center-of-mass energy $\sqrt{s}$, ranging from $0.35$~TeV to $3$~TeV, covering the energy reach of future $e^+e^-$ colliders such as the FCC-ee, ILC, and CLIC. The results are shown for several values of the NC scale $\Lambda_{\mathrm{NC}}$ and for both transverse and longitudinal space-time deformations. Throughout, we use $m_t=173$~GeV.

In contrast to purely space-space noncommutativity, which typically suppresses the cross section in other processes, space-time noncommutativity leads to an enhancement that grows with $\sqrt{s}$. This enhancement becomes more pronounced for smaller values of $\Lambda_{\mathrm{NC}}$, and is particularly strong for longitudinal deformations with $\boldsymbol{\xi}=(0,0,1)$. 

Higher center-of-mass energies therefore provide an effective probe of TeV-scale space-time NC deformations, where modest deviations from the SM prediction may signal physics beyond the SM. For example, at $\sqrt{s}=3$~TeV the SM prediction for
$\sigma(e^+e^-\to t\bar{t})$ is approximately $0.018$~pb. A transverse NC deformation shifts this value to about $0.022$~pb, corresponding to an enhancement of roughly $23\%$ for $\Lambda_{\mathrm{NC}}=1.6$~TeV, while longitudinal deformations can lead to enhancements exceeding $100\%$ at the same scale. Consequently, precision measurements at future $e^+e^-$ colliders could realistically exclude NC scales in the TeV range and potentially extend sensitivity into the multi-TeV regime. Similar conclusions were reached in ref.~\cite{Das:2012hpe} for Higgs-pair production in the NCSM.

\begin{figure}[ht]
\centerline{\includegraphics[width=.45\textwidth]{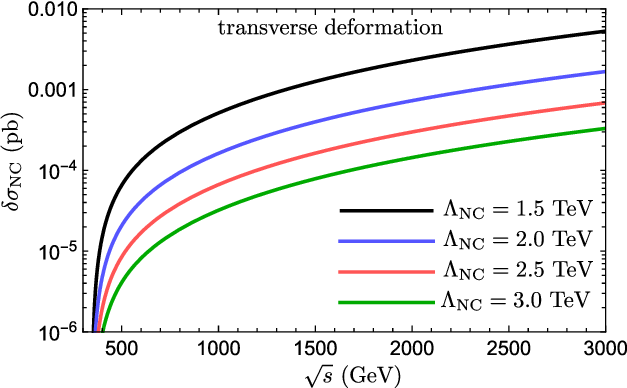}
\includegraphics[width=.45\textwidth]{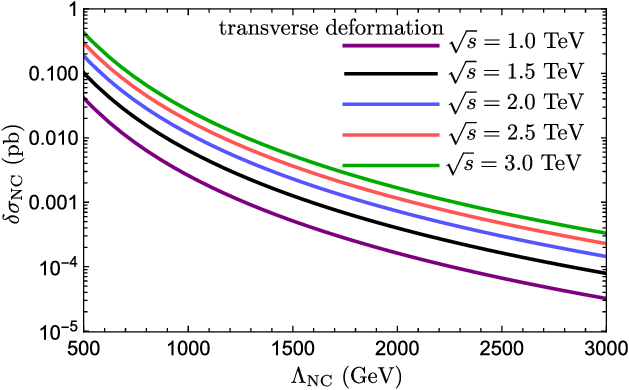}}
\vspace*{8pt}
\caption{\label{fig:delsiig}NC correction to the total cross-section for transverse space-time deformations as a function of the center-of-mass energy $\sqrt{s}$ (left) and of the NC scale $\Lambda_{\mathrm{NC}}$ (right).\protect}
\end{figure}

To further quantify NC effects, we show in figure.~\ref{fig:delsiig} the NC correction $\delta\sigma_{\mathrm{NC}}$ as a function of $\sqrt{s}$ for different values of $\Lambda_{\mathrm{NC}}$, and as a function of $\Lambda_{\mathrm{NC}}$ for fixed $\sqrt{s}$, in the case of transverse deformations.
These results confirm that NC effects decrease with increasing $\Lambda_{\mathrm{NC}}$ and grow with $\sqrt{s}$, highlighting the importance of high-energy collisions for probing NC geometry.

\subsubsection{Collision threshold energy}

We adopt a \emph{detection-threshold} strategy and introduce a benchmark collision energy $\sqrt{s_0}$, defined as the minimum machine centre-of-mass energy required for the NC contribution to the cross section, $\delta\sigma_{\mathrm{NC}}$, to reach a fixed fraction $p$ of the corresponding SM prediction. Explicitly, $\sqrt{s_0}$ is determined by the condition
\begin{equation}\label{eq:thrsh}
\frac{\delta\sigma_{\mathrm{NC}}(\sqrt{s_0})}{\sigma_{\mathrm{SM}}(\sqrt{s_0})} = p\,,
\end{equation}
where we take $p=10\%$ for illustrative purposes. This choice represents a conservative benchmark corresponding to a measurable deviation that could plausibly yield a $5\sigma$ indication of NC effects in precision cross-section measurements.

\begin{figure}[ht]
\centerline{\includegraphics[width=.55\textwidth]{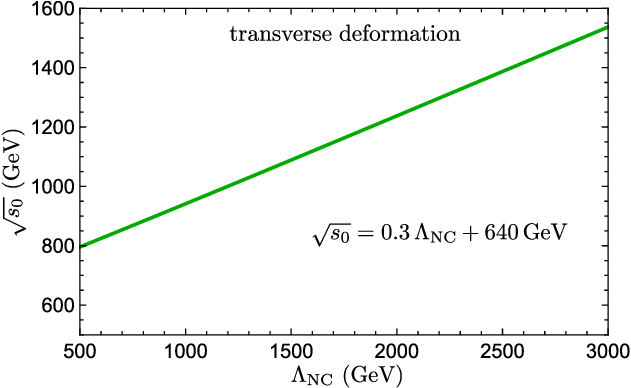}}
\vspace*{8pt}
\caption{\label{fig:s0}Threshold centre-of-mass energy $\sqrt{s_0}$ as a function of the NC scale $\Lambda_{\mathrm{NC}}$ for transverse space-time deformations.\protect}
\end{figure}

For a fixed value of the NC scale $\Lambda_{\mathrm{NC}}$, the threshold energy $\sqrt{s_0}$ is obtained by numerically solving eq. \eqref{eq:thrsh}. The resulting values of $\sqrt{s_0}$ as a function of $\Lambda_{\mathrm{NC}}$ are displayed in figure \ref{fig:s0}. The numerical data exhibit an approximately linear dependence between the two quantities over the range of parameters considered.

To quantify this behaviour, we perform a linear interpolation of the numerical results shown in figure \ref{fig:s0}. This yields the following empirical parametrization for the threshold energy,
\begin{equation}\label{eq:s0}
\sqrt{s_0}=\mathcal{A}\,\Lambda_{\mathrm{NC}}+\mathcal{B}\,,\qquad \mathcal{A}= 0.3\,,\qquad\mathcal B= 640\,\mathrm{GeV}\,.
\end{equation}

This result indicates that, as the NC scale increases, the required minimum centre-of-mass energy grows more slowly, at a rate of roughly $30\%$ of the growth of $\Lambda_{\mathrm{NC}}$. Consequently, relatively modest increases in collider energy are sufficient to probe significantly larger values of the NC scale.

The primary significance of this parametrization lies in its direct applicability as a predictive tool for future experimental programs. For example, if theoretical considerations point to an NC scale of order $\Lambda_{\mathrm{NC}} \simeq 3.0~\mathrm{TeV}$, our relation predicts a minimum centre-of-mass energy of approximately $\sqrt{s_0} \simeq 1.5~\mathrm{TeV}$ to reach the target sensitivity. Conversely, the absence of an observed deviation at a given collider energy can be immediately translated into a lower bound on $\Lambda_{\mathrm{NC}}$. In this way, eq \eqref{eq:s0} provides a transparent framework for both discovery-oriented searches and the interpretation of null results, hence offering clear guidance for future investigations of NC effects in the top-quark sector.

\subsection{Polar distribution and forward-backward asymmetry}

\begin{figure}[ht]
\centerline{\includegraphics[width=.45\textwidth]{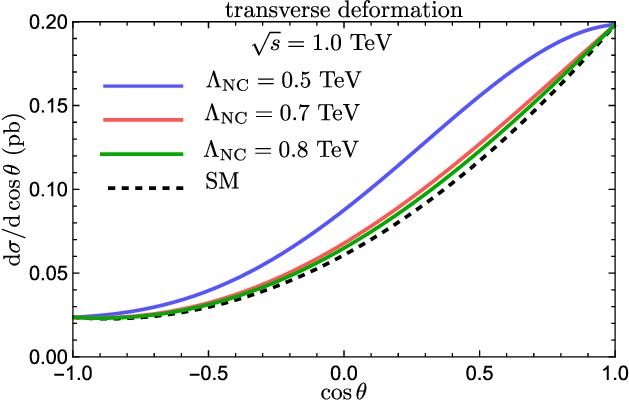}
\includegraphics[width=.45\textwidth]{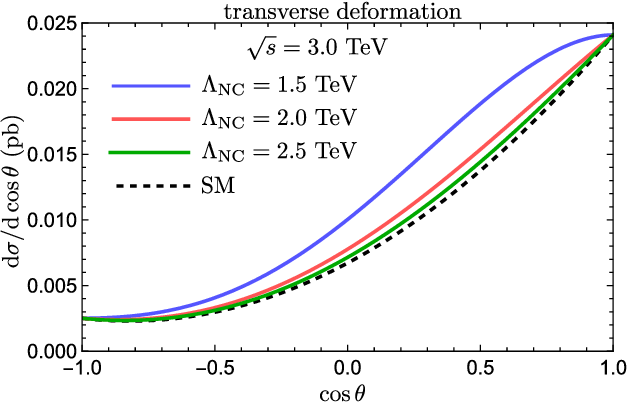}}
\vspace*{8pt}
\centerline{\includegraphics[width=.45\textwidth]{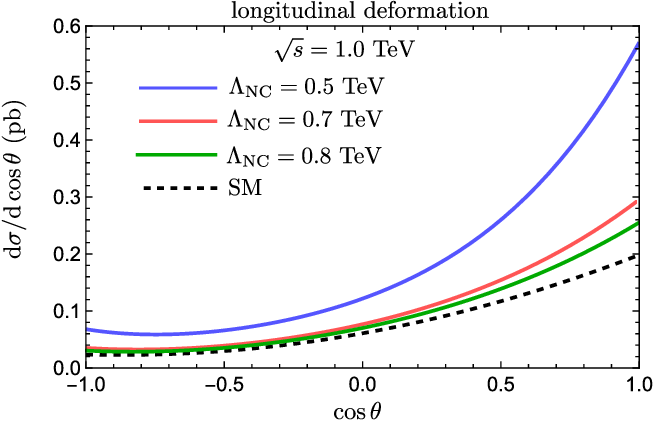}
\includegraphics[width=.45\textwidth]{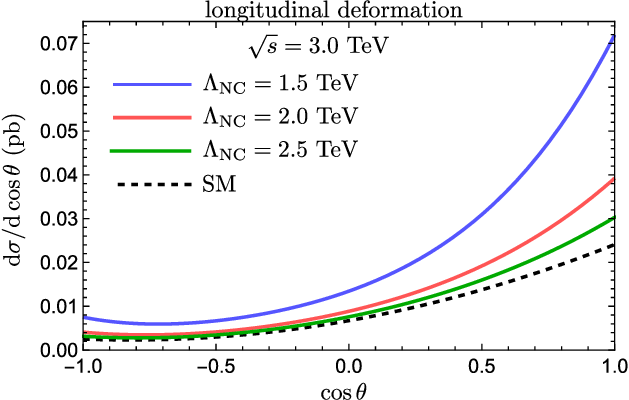}}
\vspace*{8pt}
\caption{\label{fig:polar}Differential polar-angle distribution of the top quark for transverse (top panels) and longitudinal (bottom panels) NC deformations.\protect}
\end{figure}

Turning to differential observables, we now analyze the polar-angle distribution of the top quark together with the associated forward-backward asymmetry. As in the case of the total cross section, azimuthal averaging decomposes the differential distribution into longitudinal and transverse components, as given in eq.~\eqref{eq:azim}. Employing this azimuthal-averaged NC correction, we compute and display the differential polar distribution $\d\sigma/\d\cos\theta$ as a function of $\cos\theta$ in figure \ref{fig:polar}. The results are shown for several representative values of the NC scale $\Lambda_{\mathrm{NC}}$, at centre-of-mass energies $\sqrt{s}=1~\mathrm{TeV}$ and $\sqrt{s}=3~\mathrm{TeV}$, and for both transverse and longitudinal deformations of space-time noncommutativity. The chosen values of $\Lambda_{\mathrm{NC}}$ are taken sufficiently large compared to $\sqrt{s}$ in order to avoid the unitarity issues discussed previously.

As can be seen from the figure, the polar-angle distributions exhibit a pronounced sensitivity to the NC scale when $\Lambda_{\mathrm{NC}}$ is of the same order as the collider energy $\sqrt{s}$. For example, at CLIC energies ($\sqrt{s}=3~\mathrm{TeV}$), a longitudinal NC deformation with $\Lambda_{\mathrm{NC}}=2.5~\mathrm{TeV}$ induces corrections at the level of $\sim 25\%$ in the forward region, $\cos\theta \simeq 1$. A comparable modification, of order $\sim 28\%$, is obtained at ILC energies ($\sqrt{s}=1~\mathrm{TeV}$) for a smaller NC scale, $\Lambda_{\mathrm{NC}}=0.8~\mathrm{TeV}$. In contrast, transverse deformations primarily affect the central region of the distribution around $\cos\theta \simeq 0$, where the induced deviations are more modest, reaching the level of approximately $6$-$7\%$ for the said energy scales.

The asymmetry of the polar-angle distribution can be quantified more precisely by the top-quark forward-backward asymmetry, defined as
\begin{equation}
A_{\mathrm{FB}}^{t}=\frac{\sigma_{\mathrm{forward}}-\sigma_{\mathrm{backward}}}{\sigma_{\mathrm{forward}}+\sigma_{\mathrm{backward}}}\,,
\end{equation}
where $\sigma_{\mathrm{forward}}$ [$\sigma_{\mathrm{backward}}$] is obtained by integrating the differential distribution over the forward ($0<\cos\theta<1$) [backward ($-1<\cos\theta<0$)] hemisphere. This observable is well known for its sensitivity to deviations from the SM and is widely employed in experimental analyses as a sensitive probe in searches for new physics scenarios.

\begin{figure}[ht]
\centerline{\includegraphics[width=.47\textwidth]{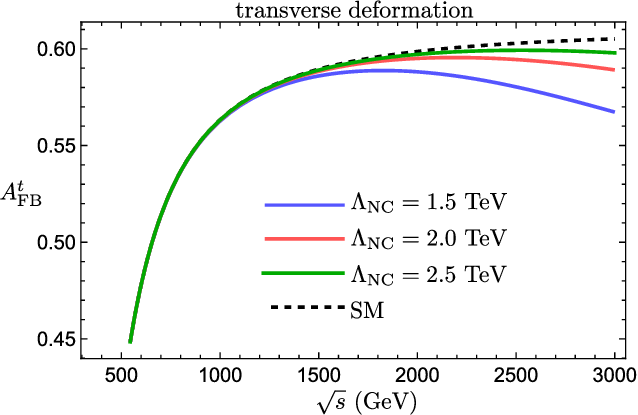}
\includegraphics[width=.47\textwidth]{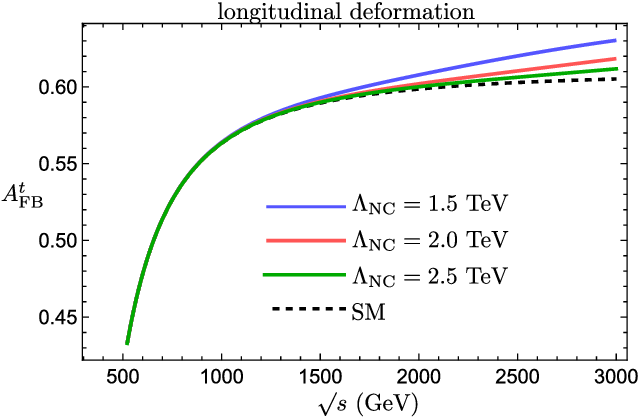}}
\vspace*{8pt}
\caption{\label{fig:FAB}Forward-backward asymmetry of the top quark for transverse (left) and longitudinal (right) NC deformations.\protect}
\end{figure}
In figure \ref{fig:FAB} we present $A_{\mathrm{FB}}^{t}$ as a function of the centre-of-mass energy. Within the SM, the forward production of top quarks is strongly enhanced, with the asymmetry rising with $\sqrt{s}$ and reaching values of about $0.6$ at $\sqrt{s}=3~\mathrm{TeV}$, eventually saturating at approximately $0.61$ at very high energies. NC corrections have a significant impact on this behaviour: transverse deformations tend to reduce the forward-backward asymmetry, while longitudinal deformations enhance it.

Quantitatively, at $\sqrt{s}=3~\mathrm{TeV}$ and for $\Lambda_{\mathrm{NC}}=2.5~\mathrm{TeV}$, we find a downward (upward) shift of the asymmetry at the level of $\sim 1\%$ for transverse (longitudinal) deformations. More generally, a collider operating at this energy would be sensitive to NC scales of order $\Lambda_{\mathrm{NC}}\sim 1.5~\mathrm{TeV}$, provided experimental uncertainties can be controlled at the $5\%$ level, corresponding to deviations of this magnitude from the SM prediction. The experimental capabilities of a machine such as CLIC--including its high integrated luminosity at the inverse-attobarn level, excellent detector performance, and clean event reconstruction--make precision measurements of this kind feasible. Such measurements could either impose stringent constraints on the NC scale or provide the first hints of its possible existence in the top-quark sector.

\subsection{Azimuthal modulation}

Within the SM, the azimuthal-angle distribution of the produced top quark is strictly flat, reflecting the rotational symmetry of the underlying dynamics around the beam axis. In contrast, NC effects introduce additional momentum-dependent structures in the squared matrix element, thus breaking this isotropy. As a result, the azimuthal distribution constitutes a particularly sensitive observable for probing space-time noncommutativity and has been widely exploited in phenomenological studies of this kind.

In the case of space-time noncommutativity, the NC correction factor $\mathcal{N}$ acquires an explicit dependence on the azimuthal angle through the transverse components of the deformation vector $\boldsymbol \xi$. Consequently, purely longitudinal deformations lead only to an overall rescaling of the azimuthal distribution, leaving its shape unchanged. By contrast, for purely transverse deformations, the functional form of $\mathcal{N}$ implies a characteristic sinusoidal modulation of the azimuthal distribution, with an amplitude and phase governed by the deformation factor given in eq.~\eqref{eq:Nt}.

\begin{figure}[ht]
\centerline{\includegraphics[width=.45\textwidth]{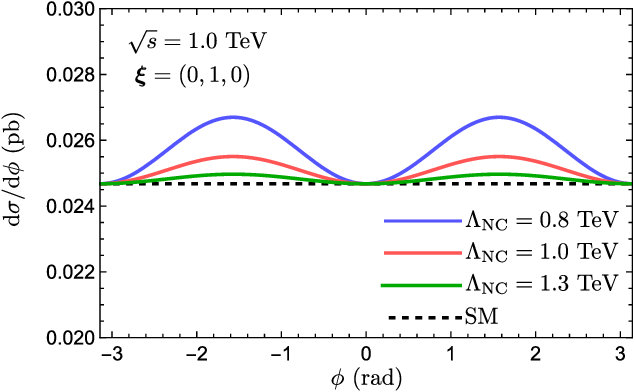}
\includegraphics[width=.45\textwidth]{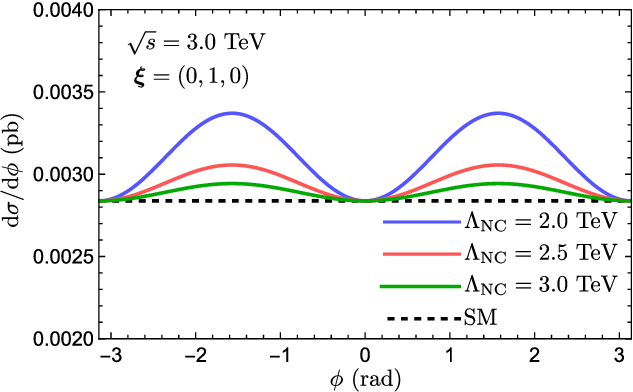}}
\vspace*{8pt}
\centerline{\includegraphics[width=.45\textwidth]{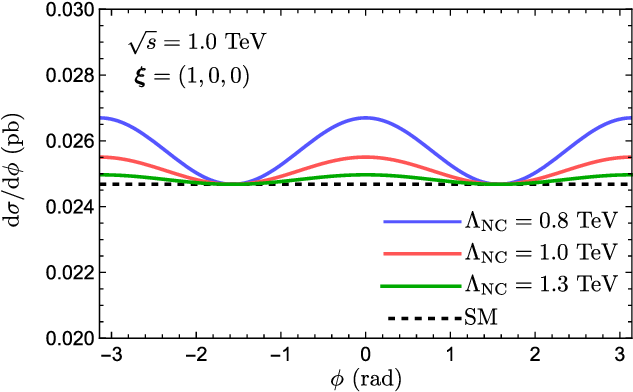}
\includegraphics[width=.45\textwidth]{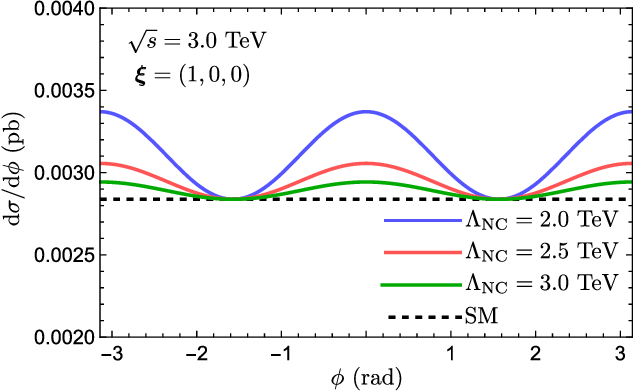}}
\vspace*{8pt}
\caption{\label{fig:azimuthal}Differential azimuthal-angle distribution for transverse deformations along the $y$ axis (top) and the $x$ axis (bottom).\protect}
\end{figure} 
In figure \ref{fig:azimuthal}, we present the differential azimuthal distribution $\d\sigma/\d\phi$ as a function of the azimuthal angle $\phi$ at centre-of-mass energies $\sqrt{s}=1~\mathrm{TeV}$ and $\sqrt{s}=3~\mathrm{TeV}$, for transverse deformations oriented along the $x$ and $y$ axes. Deformations in the transverse plane induce a smooth phase shift when moving from an $x$-axis to a $y$-axis deformation, directly reflecting the sinusoidal structure of the NC correction factor in Eq.~\eqref{eq:Nt}.

The impact of space-time noncommutativity on this distribution is particularly pronounced at higher energies. Focusing on $\sqrt{s}=3~\mathrm{TeV}$, we find that for a NC scale of $\Lambda_{\mathrm{NC}}=3~\mathrm{TeV}$ the azimuthal distribution is distorted by up to approximately $4\%$ at its extrema, while for $\Lambda_{\mathrm{NC}}=2.5~\mathrm{TeV}$ the modification increases to about $7\%$. Assuming a conservative experimental sensitivity at the $5\%$ level for measurements of this observable at future $e^+e^-$ colliders, these results indicate that the azimuthal distribution could provide a powerful handle for either detecting space-time noncommutative effects or setting meaningful lower bounds on the NC scale at the level of a few TeV.

\section{Conclusions}

In this work we have investigated the phenomenological consequences of space-time noncommutativity in the process $e^+e^- \to t\bar t$ within the framework of the mNCSM. Focusing on both total and differential observables, we have analyzed how NC effects modify the production rate and angular distributions of top-quark pairs at high-energy lepton colliders.

We first established the kinematical setup and clarified the role played by different choices of the NC deformation tensor. While purely space-space noncommutativity leads to vanishing leading-order corrections for the $s$-channel process under consideration, allowing for space-time noncommutativity generates nontrivial and potentially observable effects. Although such deformations are known to induce apparent unitarity-violating behavior at very high energies, we have emphasized that this should be interpreted as a limitation of the effective theory description. Accordingly, our analysis was restricted to energy regimes $\sqrt{s}\lesssim \Lambda_{\mathrm{NC}}$, where differential observables remain under perturbative control and provide reliable probes of NC dynamics.

Using a benchmark ``detection threshold'' criterion, we quantified the sensitivity of future $e^+e^-$ colliders to the NC scale through total cross-section measurements. We found an approximately linear relation between the threshold center-of-mass energy and the NC scale, which offers a simple and practical guideline for assessing the discovery reach or exclusion potential of a given collider setup. This relation allows null results at a fixed energy to be directly translated into lower bounds on $\Lambda_{\mathrm{NC}}$.

We then turned to differential observables, which constitute the most robust signatures of noncommutativity. The polar-angle distribution of the top quark and the associated forward-backward asymmetry were shown to be sensitive to both transverse and longitudinal deformations of space-time. In particular, longitudinal deformations enhance the forward-backward asymmetry, while transverse deformations tend to reduce it. At multi-TeV center-of-mass energies, percent-level deviations from the SM predictions were found for NC scales in the TeV range, indicating that precision measurements of these observables could either reveal hints of noncommutativity or significantly tighten existing bounds.

Finally, we studied the azimuthal distribution of the top quark, which is flat in the SM but acquires characteristic sinusoidal modulations in the presence of transverse space-time noncommutativity. We showed that these azimuthal modulations can reach several percent at $\sqrt{s}=3$~TeV for NC scales of $\Lambda_{\mathrm{NC}}\sim 2.5$--$3$ TeV, making them particularly clean and sensitive probes of noncommutative effects. Given the clean experimental environment, high luminosity, and excellent reconstruction capabilities anticipated at future linear colliders, such measurements appear well within realistic experimental reach.

In summary, our results demonstrate that top-quark pair production at high-energy $e^+e^-$ colliders provides a powerful and complementary avenue for probing space-time noncommutativity. Differential observables, especially polar and azimuthal angular distributions, emerge as particularly sensitive and theoretically robust probes. Future precision measurements in the top-quark sector therefore have strong potential either to uncover signatures of noncommutative space-time or to push the corresponding NC scale well into the multi-TeV regime.

\section*{Acknowledgments}

This work was supported by PRFU research project No. B00L02UN050120230003. The authors gratefully acknowledge financial support from the Algerian Ministry of Higher Education and Scientific Research and the Directorate General for Scientific Research and Technological Development (DGRSDT).

\appendix

\section{Feynman rules}\label{sec:Feyn}

Following the formulation of the mNCSM presented in ref.~\cite{Melic:2005am}, we summarize here the NC Feynman rules relevant for the present work. In particular, we list the vertices involving a fermion-antifermion pair of mass $m_f$ and a neutral gauge boson, $ff\gamma$ and $ffZ$, including corrections up to first order in the noncommutativity parameter $\Theta^{\mu\nu}$, i.e.\ $\mathcal{O}(\Theta)$.

\begin{figure}[ht]
\centerline{\includegraphics[width=.25\textwidth]{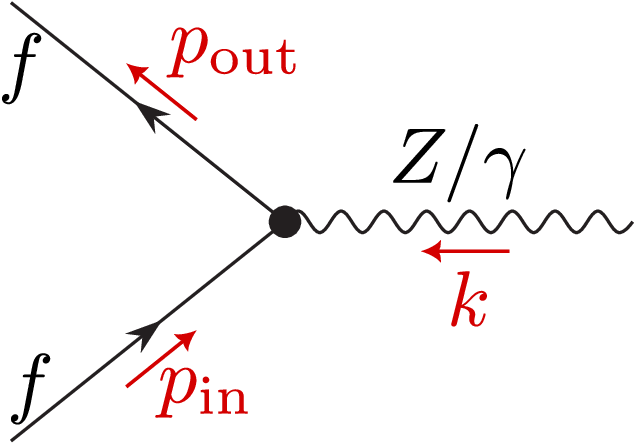}}
\vspace*{8pt}
\caption{The $ff\gamma/Z$ vertex in the mNCSM. The fermion momentum $p_{\mathrm{in}}$ flows into the vertex and $p_{\mathrm{out}}$ flows out, while the gauge-boson momentum $k$ is taken to flow into the vertex.\protect\label{Vert}}
\end{figure}

The NC Feynman rule for the vertex $f(p_{\mathrm{in}})\text{--}f(p_{\mathrm{out}})\text{--}\gamma/Z(k)$, depicted in figure \ref{Vert}, is given by \cite{Melic:2005am}
\begin{subequations}
\begin{align}
\mathcal{V}_{ff\gamma}&=ieQ_f\bigg[\gamma_\mu\left(1-\frac{i}{2}(p_{\mathrm{out}}\Theta p_{\mathrm{in}})\right)\notag\\
&+\frac{i}{2}(p_{\mathrm{out}}\Theta)_\mu\,(\slashed{p}_{\mathrm{in}}-m_f)+\frac{i}{2}(\Theta p_{\mathrm{in}})_\mu\,(\slashed{p}_{\text{out}}-m_f) \bigg],\\
\mathcal{V}_{ffZ}&=\frac{ie}{\sin (2\theta_w)}\bigg[\gamma_\mu\left(c_V^f-c_A^f\,\gamma_5\right)\left(1-\frac{i}{2}(p_{\mathrm{out}}\Theta p_{\mathrm{in}})\right)\notag\\
&+\frac{i}{2}(p_{\mathrm{out}}\Theta)_\mu\left(c_V^f+c_A^f\,\gamma_5\right)(\slashed{p}_{\mathrm{in}}-m_f)+\frac{i}{2}(\Theta p_{\mathrm{in}})_\mu\,(\slashed{p}_{\mathrm{out}}-m_f)\left(c_V^f-c_A^f\,\gamma_5\right)\bigg],
\end{align}
\end{subequations}
where $(\Theta p_i)_\mu\equiv \Theta_{\mu\nu}p_i^\nu=-(p_i\Theta)_\mu$. Here $p_{\mathrm{in}}$ denotes the momentum of the fermion line entering the vertex, while $p_{\mathrm{out}}$ corresponds to the momentum of the fermion line leaving the vertex. Fermion momentum flow follows the direction of the fermion line, and the gauge-boson momentum $k$ is always taken to flow into the vertex. Momentum conservation at the vertex implies $p_{\mathrm{in}}+k=p_{\mathrm{out}}$.

For the $e^+e^-\gamma/Z$ vertex appearing in figure \ref{fig1}, the momenta are identified as $p_{\mathrm{in}}=p_1$, $p_{\mathrm{out}}=-p_2$, and $k=-p_1-p_2$. The corresponding vertex factor is sandwiched between Dirac spinors as $\bar{v}(p_2)\,\mathcal{V}_{ee\gamma/Z}\,u(p_1)$. Making use of the equations of motion,
\begin{equation}
\bar{v}(p_2)(\slashed{p}_2+m_e)=0\,, \qquad(\slashed{p}_1-m_e)u(p_1)=0\,,
\end{equation}
effectively allows us to  replace $-\slashed{p}_2-m_e \to 0$ and $\slashed{p}_1-m_e \to 0$ both in the $\gamma$ and $Z$ vertices. One thus finds that the terms proportional to $(\slashed{p}_{\mathrm{in}}-m_f)$ and $(\slashed{p}_{\mathrm{out}}-m_f)$ vanish identically for both the photon and $Z$ vertices.

An analogous simplification applies to the top-quark vertex, where $p_{\mathrm{in}}=-p_4$ and $p_{\mathrm{out}}=p_3$, and the vertex is inserted between spinors as $\bar{u}(p_3)\,\mathcal{V}_{tt\gamma/Z}\,v(p_4)$. Using
\begin{equation}
(\slashed{p}_4+m_t)v(p_4)=0\,, \qquad\bar{u}(p_3)(\slashed{p}_3-m_t)=0\,,
\end{equation}
the momentum-dependent terms of the vertices $(\slashed{p}_{\mathrm{in}}-m_f)$ and $(\slashed{p}_{\mathrm{out}}-m_f)$ again vanish. Consequently, only the leading $\gamma_\mu$ terms in each vertex contribute to the amplitudes, as indicated in Eq.~\eqref{eq:amplitudes}.

\bibliographystyle{ws-mpla}
\bibliography{ref}

\end{document}